\documentclass[prb,twocolumn,showpacs,showkeys]{revtex4}
\usepackage{graphicx,epsfig}
\usepackage{bm}
\usepackage{dcolumn}

\def\bea{\begin{eqnarray}}
\def\eea{\end{eqnarray}}
\def\be{\begin{equation}}
\def\ee{\end{equation}}
\def\pp{\partial}

\begin{document}

\title{The electron-phonon coupling is large for localized states }

\author{Raymond Atta-Fynn}
\email{attafynn@phy.ohiou.edu}
\author{Parthapratim Biswas}
\email{biswas@phy.ohiou.edu}
\author{D.~A.~Drabold}
\email{drabold@ohio.edu}

\affiliation{Department of Physics and Astronomy, Ohio University, Athens, OH 45701}



\begin{abstract}
From density functional calculations, we show that localized states stemming 
from defects or topological disorder exhibit an anomalously large electron-phonon 
coupling. We provide a simple analysis to explain the observation and perform a 
detailed study on an interesting system: amorphous silicon. We compute first 
principles deformation potentials  (by computing the sensitivity of specific 
electronic eigenstates to individual classical normal modes of vibration). We 
also probe thermal fluctuations in electronic eigenvalues by first principles 
thermal simulation. We find a strong correlation between a static property of the 
network [localization, as gauged by inverse participation ratio (IPR)] and a 
dynamical property (the amplitude of thermal fluctuations of electron energy 
eigenvalues) for localized electron  states.  In particular, both the electron-phonon 
coupling and the variance of energy eigenvalues are proportional to the IPR 
of the localized state. We compare the results for amorphous Si to photoemission 
experiments. While the computations are carried out for silicon, very similar 
effects have been seen in other systems with disorder.  
\end{abstract}

\maketitle
\section{INTRODUCTION}

Electron states of finite spatial extent, so called ``localized" states are ubiquitous 
in nature. Localized molecular orbitals occur in molecules and in solid state 
systems with disorder (this could take the form of defects, such as a dangling 
bond state or more subtly from topological and or chemical disorder in amorphous 
materials or glasses). Such localized states are also known in polymers. Surface 
states, as another form of defect, also are often spatially compact. The study of 
localization in its own right has been an important and active subfield of 
condensed matter theory since the fifties.

Another physical quantity of key importance is the electron-phonon (e-p) coupling, 
the interaction term connecting the electronic and lattice systems. Perhaps most 
spectacularly, the e-p coupling is the origin of superconductivity as expressed 
in BCS theory\cite{Bardeen}.\,\,Phillips\cite{Phillips} has shown that large e-p couplings 
in the cuprate superconductors can lead to a successful model of high T$_c$ 
superconductivity\cite{Bednorz} within the framework of conventional BCS 
superconductivity. The e-p coupling is also the mediator of all light-induced 
structural changes in materials. In amorphous silicon the greatest outstanding 
problem of the material, the Staebler-Wronski effect\cite{Staebler}, depends 
critically upon the electron-lattice interaction. A zoo of analogous effects is 
studied in glasses; perhaps the most important example is reversible photo-amorphization 
and photo-crystallization used in the GeSbTe phase-change materials used in 
current writable CD and DVD technology.

Previous thermal simulations with Bohn-Oppenheimer dynamics have indicated 
that there exists a large electron-phonon coupling for the localized 
states in the band tails and in the optical gap~\cite{Li,Drabold1}. 
Earlier works on chalcogenide glasses by Cobb and Drabold~\cite{Cobb} have 
emphasized a strong correlation between the thermal fluctuations as gauged 
by root mean square (RMS) variation in the LDA eigenvalues and wave function 
localization of a gap or tail state (measured by inverse participation 
ratio\cite{Drabold2}, a simple measure of localization). Drabold and 
Fedders\cite{Drabold3} have also shown that localized eigenvectors may 
fluctuate dramatically even at room temperature. Recently, Li and 
Drabold relaxed the adiabatic (Born-Oppenheimer) approximation to 
track the time-development of electron packets scattered by lattice vibrations\cite{Li}.
In this paper, we examine the electron phonon coupling and provide a heuristic 
analysis of the e-p coupling for localized electron states. We explore the e-p 
coupling in some detail for a particular model system (amorphous silicon) which 
provides us with a convenient variety of localized, partly localized ``bandtail" 
and extended states. We compute  deformation potential (which measures the response 
of a selected electron state to a particular phonon), and also track thermally-induced 
fluctuations of electronic eigenvalues. We find that localized states always exhibit 
a large e-p coupling. Our computations are carried out using a first principles 
molecular dynamics code {\sc Siesta}, and the eigenvalues and states that we 
study are from the Kohn-Sham equations with a rich local orbital basis. A rationale 
for the study of the Kohn-Sham states is given elsewhere\cite{Attafynn}. We 
emphasize that the results that we give are qualitatively general -- not just 
an artifact of studying a disordered phase of silicon (we have, for example, 
seen exactly the same effects in various binary glasses which exhibit very 
different topological and chemical disorder).
\section{THEORY}

To establish a connection between electron-phonon coupling and 
wave function localization for the electrons, we consider an 
electronic eigenvalue $\lambda_n$ near the band gap of a-Si. The 
sensitivity of $\lambda_n$ due to an arbitrary small displacement 
of an atom (possibly thermally induced) can be estimated using the 
Hellmann-Feynman theorem~\cite{Feynman},
\[
\frac{\pp \lambda_n}{\pp \bf R_{\alpha}} = \langle 
\psi_n|{\frac{\pp \bf H}{\pp \bf R_{\alpha}}}|\psi_n \rangle.
\] 
Here we have assumed that the basis functions are fixed and 
$|\psi_n \rangle$ are the eigenvectors of the Hamiltonian {\bf H}. 
For small lattice distortion $\{\delta\bf R_\alpha\}$, the 
corresponding change in $\delta\lambda_n$ is, 
\be
\label{first}
\delta\lambda_n \approx \sum_{\alpha = 1}^{3N}\,\langle \psi_n|\frac{\partial {\bf H}}{\partial {\bf 
R_{\alpha}}}|\psi_n \rangle\,\delta\bf R_\alpha
\ee
where $\rm N$ is the total number of atoms in the model. If the displacement 
$\delta\bf R_{\alpha}(\rm t)$  arises from classical vibrations, one can 
write~\cite{Note1}, 
\be
\label{second}
\delta \bf R_{\alpha}(\rm t) = \sum_{\omega = 1} ^{3N} \rm A(\rm T,\omega)\,\cos(\omega\,t 
+ \phi_\omega)\,\chi_{\alpha}(\omega),
\ee
where $\rm \omega$ indexes the normal mode frequencies, $\rm A(\rm T,\rm \omega)$ is 
the temperature dependent amplitude of the mode with frequency $\rm \omega$,  
$\phi_\omega$ is an arbitrary phase, $ \rm \chi_{\rm \alpha}(\rm \omega)$ is a normal mode 
with frequency  $\omega$ and vibrational displacement index $\alpha$. Using
the temperature dependent squared amplitude
$ \rm A^2 = \rm 3k_BT/\rm M \rm \omega^2$, 
the trajectory (long time) average of $\rm \delta \rm\lambda_{n}^2$ 
can be written (using Eq.\,\ref{first} and \ref{second}) as, 

\be
\label{eq-1}
\langle {\delta \lambda_{n}^2}\rangle  = \lim_{\tau \rightarrow \infty}\frac{1}{\tau}\int_0^\tau dt \, \delta 
\lambda^2_n(t) \approx \rm \left(\frac{3k_BT}{2M}\right) \sum_{\omega=1}^{3N}\,\frac{\Xi_{n}^2(\omega)}{\omega^2},
\ee
where the electron-phonon coupling $\rm \Xi_{n}(\omega)$ is given by,

\be
\label{eq-2}
\Xi_{n}(\omega) = \sum_{\alpha = 1}^{3N} \langle \psi_n|\frac{\partial{\bf H}}{\partial{\bf R_{\alpha}}}|\psi_n
\rangle\,\chi_{\alpha}(\omega).
\ee
One can infer from Eq.\,\ref{eq-1} that thermally induced fluctuation
in the energy eigenvalues is a consequence of electron-phonon 
coupling.  Note that for a given electronic eigenvalue, the 
contribution to the coupling comes from the entire vibrational spectrum 
involving all the atoms in the systems. Since the normalized eigenstate
can be written as $|\psi_n\rangle = \sum_{i}a_{ni}\,|i\rangle$ 
where $|i\rangle$ are the basis orbitals, it follows from Eq.\,\ref{eq-2} that, 

\begin{widetext}
\bea
\label{eq-3}
\Xi^2_{n}(\omega) &=& \sum_{\alpha,\beta,i,j,k,l}\,
a^{*}_{ni}a_{nj}a^{*}_{nk}a_{nl}\langle i|\frac{\partial \bf H}{\partial \bf R_{\alpha}}|j 
\rangle\,\langle k|\frac{\partial \bf H}{\partial \bf R_{\beta}}|l\rangle \, \chi_{\alpha}(\omega) 
\, \chi_{\beta}(\omega) \nonumber \\ 
&=& \sum_{i,\alpha}|a_{ni}|^4\, [\langle i|\frac{\partial \bf H}
{\partial \bf R_{\alpha}}|i \rangle]^2  \, \chi_{\alpha}^2(\omega) + 
\sum_{ijkl\alpha \beta}^{\prime} a^{*}_{ni}a_{nj}a^{*}_{nk}a_{nl}\langle i|\frac{\partial \bf H}
{\partial \bf R_{\alpha}}|j \rangle\,\langle
 k|\frac{\partial \bf H}{\partial \bf R_{\beta}}|l\rangle 
\, \chi_{\alpha}(\omega) \,\chi_{\beta}(\omega)
\eea
\end{widetext}

The first term in the second line of Eq.\,\ref{eq-3} is positive definite (diagonal) 
while the second one, the off-diagonal term (indicated by the prime), is not of a single sign. 
In the event that only a few $a_{ni}$ dominate (the case for localized states), then the 
leading contribution to the electron-phonon coupling
originates largely from the diagonal term. The addition of a large number 
of terms of mixed sign and small magnitude leads to cancellations in the off-diagonal term leaving behind a small contribution to electron-phonon 
coupling. By comparing to direct calculations with the full Eq.\,\ref{eq-3}, we show 
that dropping the second term appears to be reasonable for well-localized electron 
states.  The approximate ``diagonal" electron-phonon coupling can be written as, 

\bea
\label{eq-4}
\Xi^2_{n}(\omega) & \approx & \sum_{\alpha  = 1}^{3N}\sum_{i=1}^{N_b} \,|a_{ni}|^4\,
[\langle i|\frac{\partial{\bf H}}{\partial{\bf R_{\alpha}}}|i \rangle]^2 
\chi_{\alpha}^2(\omega) \nonumber \\ 
&  =  & \sum_{\alpha = 1}^{3N} \sum_{i=1}^{N_b}\, {q^2_{ni}}\, 
[\langle i|\frac{\partial{\bf  H}}{\partial{\bf R_{\alpha}}}|i \rangle]^2  
\chi_{\alpha}^2(\omega)
\eea

where N$_b$ is the number of basis orbitals and $q_{ni} = |a_{n,i}|^2$ is 
the charge sitting on the $i$th orbital for a given normalized eigenstate 
$|\psi_n \rangle$. The degree of wave function localization can be measured
by defining inverse participation ratio $ \rm {\cal I}$ for the eigenstates
$|\psi_n \rangle$,
\be
\label{eq-5}
\rm {\cal I}(n) = \sum_{i=1}^{\rm N_{b}}\,q^2_{ni}.
\ee
Equation \ref{eq-4} leads to an approximate but analytic connection between 
${\cal I}$ and electron-phonon coupling. Since $\rm {\cal I}$ is large for 
localized states, one expects $\Xi_{n}(\omega)$ (and therefore $\langle 
\delta \lambda^2_n \rangle$) to be large for a localized state. If we further
assume that $\gamma^2(\omega,i)= \sum_{\alpha = 1}^{3N}
[\langle i|\partial{\bf  H}/\partial{\bf R_{\alpha}}|i \rangle]^2  
\chi_{\alpha}^2(\omega)$ is weakly dependent upon site/orbital index $i$, 
then 

\be
\label{third} 
\Xi^2_{n}(\omega)~\sim {\cal I}_n \times f(\omega), 
\ee

where $f(\omega)$ is defined from $\gamma^2$ and Eq.\,\ref{eq-4}. In 
this ``separable" approximation, it is also the case that $\langle \delta \lambda^2_n\rangle \propto {\cal I}_n$.  

\begin{figure}
\includegraphics[width=3.5in, height=3.5in]{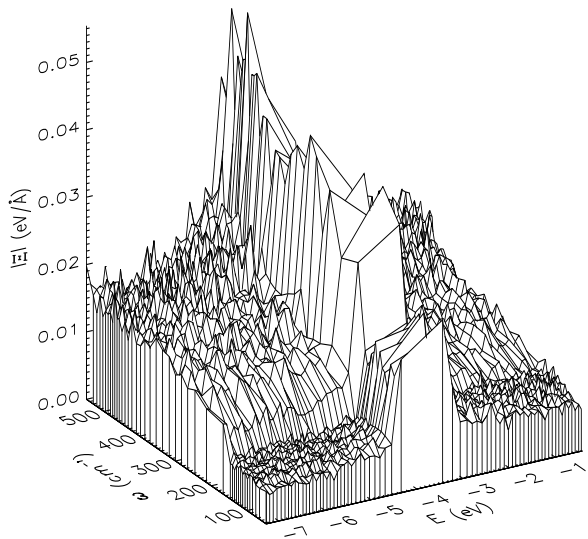}
\caption{\label{fig1}
Electron-phonon coupling surface plot for a 216-atom model 
of \emph{a}-Si. The absolute value of electron-phonon coupling 
$|\Xi|$ (cf. Eq.\,\ref{eq-2}) is plotted as a function of phonon 
frequency $\omega$ and energy eigenvalues near the gap. 
The largest value of $|\Xi|$ in the plot corresponds to the 
eigenvalue for HOMO, which is the most localized state in the 
spectrum.}
\end{figure}

\section{Method}

The model of \emph{a}-Si we have used in our calculations was
generated by Barkema and Mousseau~\cite{Barkema} using an improved
version of the Wooten, Winer and Weaire (WWW) algorithm~\cite{Wooten}.
The details of the construction was reported in Ref.\,\onlinecite{Barkema}. 
The model consists of 216 atoms of Si packed inside a cubic box of length 16.282{\AA}
and has two 3-fold coordinated atoms. The average bond angle is 
$109.5^{\circ}$ with a root mean square deviation of $11.0^{\circ}$.
The density functional calculations were performed within the local density
approximation (LDA) using the first principles code {\sc Siesta}
\cite{Siesta1,Siesta2,Siesta3}. We have used a non self-consistent version of
density functional theory based on the linearization of the Kohn-Sham
equation by Harris functional approximation~\cite{Harris} along with the
parameterization of Perdew and Zunger~\cite{Perdew} for the exchange-correlation
functional. 
%Norm conserving Troullier-Martins pseudopotentials~\cite{Troullier}
%factorized in the Kleinman-Bylander form~\cite{Kleinman} were used to
%remove the core electrons. 
The choice of an appropriate basis is found be very important and has been discussed at length 
in a recent communication~\cite{Attafynn}. While the minimal basis
consisting of one $s$ and three $p$ electrons can adequately describe the
electronic structure of amorphous silicon in general, there is some 
concern about the applicability of these minimal basis in describing deeply 
localized and low lying excited states in the conduction bands accurately. 
We have therefore employed a larger single-$\zeta$ basis with 
polarization (d) orbitals (SZP)~\cite{Sankey, Note2} in the present work. 
Throughout the calculation we have used only the $\Gamma$ point to 
sample the Brillouin zone.

\begin{figure}
\includegraphics[width=3.0in, height=3.5in]{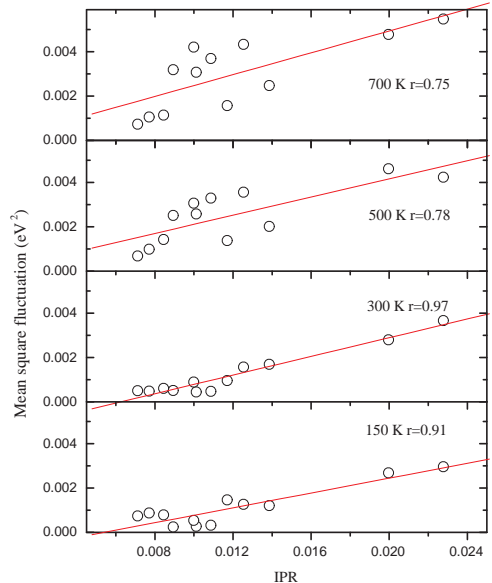}
\caption{\label{fig2} 
(Color online) Mean square fluctuations of electronic eigenvalues 
versus inverse participation ratio plot at different temperature. 
The fluctuations at temperature 150K and 300K are found to be 
linearly correlated with the participation ratio for the corresponding 
eigenstates as predicted in section II. The correlation coefficient ($r$)  
for different temperature is indicated in the plot. 
}
\end{figure}

Starting with a fully 
relaxed configuration, we construct the dynamical matrix elements by successively displacing 
each atom in the supercell along three orthogonal directions (x, y and z) 
by 0.01{\AA} and computing the forces for each configuration. 
Within the harmonic approximation, the spring constant associated with 
each atom and direction can be written as a second derivative of the total energy 
with respect to the displacement of the atom in that direction. We have checked 
the convergence of both the matrix elements by using a different set of values for 
atomic displacement and used a value of {0.01{\AA}} in our calculations. 
$\partial \lambda/\partial {\bf R}_\alpha$ was obtained by finite differences from the 
dynamical matrix calculations.  The calculation does not need any extra effort beyond that of forming the
dynamical matrix. 

To explore the validity of our analysis and to elucidate
the connection between the localization (IPR) ($\rm {\cal I}$)
of electronic eigenstates and fluctuation of the conjugate eigenvalues, we performed
thermal MD simulations at constant temperatures using a No\'se-Hoover thermostat. The 
simulations were performed at
temperatures 150K, 300K, 500K and 700K with a time step of 2.5fs for
a total period of 2.5ps. For a given temperature, the mean square fluctuations 
were computed by tracking the eigenvalues at each time step and averaging over 
the total time of simulation excepting the first few hundred time steps  to 
ensure equilibration. 
\begin{figure}
\includegraphics[width=3.5in, height=3.7in]{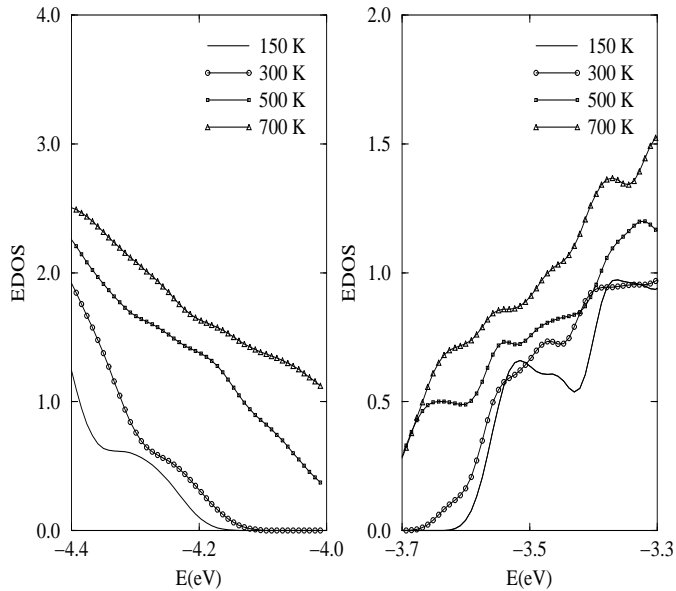}
\caption{\label{fig3} 
The average electronic density of band tails states for
four different temperature T = 150K, 300K, 500K and 700K. 
Note that the conduction band tails (right) near the Fermi level (which is between the two tails) 
are more sensitive to thermal disorder than the valence band tails (left) 
providing a qualitative agreement with experimental result in 
Ref.\,25.  
}
\end{figure}
The mean square fluctuation ($\rm {\cal R}$) for an energy 
eigenvalue $\rm \lambda_n(t)$ is defined as : 
\be
\label{eq-6}
 \rm {\cal R}_n = \langle (\lambda_n(t) - \langle \lambda_n \rangle)^2
\rangle,
\ee
where $\langle \,\rangle$ denotes the average over time.
We study the fluctuations of $\{\rm \lambda_n(\rm t)\}$ by 
plotting against time at a given temperature and compare it 
with the $\rm {\cal I}$ obtained for the corresponding 
eigenvalues. For illustrations of such adiabatic evolution of Kohn-Sham
eigenvalues, see Ref.\,~\onlinecite{Drabold3}.

\section{Results}

% Report e-p coupling obtained without approximation 

In Fig.\,\ref{fig1}, we have plotted the electron-phonon coupling 
for the states near the band gap obtained directly from Eq.\,{\ref{eq-2}. 
It is clear from the figure that the e-p coupling is large only in 
the vicinity of conduction and valence band tails. The largest e-p 
coupling in the plot corresponds to the highest occupied molecular 
orbital (HOMO) in the optical frequency regime around 415 cm$^{-1}$. 
The lowest occupied molecular orbital (LUMO) also has a large feature 
around the same frequency. 
A Mulliken charge analysis and inverse participation ratio 
calculation of the electronic eigenfunctions have shown that 
both these two states -- the HOMO and LUMO are highly localized 
and are centered around the dangling bonds present in the model. 
On moving further from the band tails in either direction 
along the energy axis, the e-p coupling drops quickly and the 
surface becomes featureless for a given eigenvalue.  This behavior of e-p coupling 
can be understood from the arguments presented in section II where we have 
shown that the e-p coupling for localized states is directly proportional 
to the inverse participation ratio. 
For a localized state, therefore, the large value of electron-phonon 
coupling can be attributed to the large value of inverse participation ratio associated 
with that state. 
Since HOMO and LUMO are the two most localized states in the spectrum, the e-p 
coupling is large for these states and as we move toward the tail states, 
the coupling decreases. It is important to note that the plot in the 
Fig.\,\ref{fig1} has been obtained from Eq.\,\ref{eq-2} without making any 
approximation and is exact inasmuch as the matrix elements obtained from 
the density functional Hamiltonian are correct. 
This observation supports our assumptions that the dominant contribution 
to e-p coupling comes from the diagonal term in Eq.\,\ref{eq-3} and 
that $\gamma^2(\omega,i)$ is weakly dependent upon site/orbital index $i$ and also indicates 
from direct simulation there exists a linear relationship between mean 
square fluctuation of electronic eigenvalues and the corresponding 
inverse participation ratio for localized states. 

%%% On the correlation between MS and IPR

In order to justify our arguments further presented in section II, we 
now give a look at the mean square fluctuation of energy eigenvalues. 
As outlined in section III, we have computed the mean square fluctuations 
at four different temperature (150K, 300K, 500K and 700K) from MD runs 
over a period of 2.5ps and plotted in the Fig.\,\ref{fig2}. The fluctuation
obtained this way provides a dynamical characteristic of the band tails 
states and is compared with a static property, the inverse participation 
ratio of the same states. 
A simple linear fit reveals a strong correlation between the eigenvalue 
fluctuation and the corresponding inverse participation ratio for the states. 
The correlation is found to be as high as $\approx$ 0.95 for T=150K and 
300K and falls to $\approx$ 0.8 at high temperature. The value of the 
correlation coefficient for different temperature is indicated in the 
Fig.\,\ref{fig2}. Once again, we see that the result is in accordance with our prediction in 
section II and provides a simple physical picture for having a large 
electron-phonon coupling for the localized states. 

%%% On thermal broadening on EDOS and Aljishi et al. experiment works.

In Fig.\,\ref{fig3}, we have plotted the time averaged electronic density 
of states for four different temperature in order to study the effect of 
thermal disorder on the tail states. It is quite clear from the figure 
that the effect of thermal broadening is quite significant on both sides of 
the gap. Photoelectron spectroscopic studies on \emph{a}-Si:H by Aljishi et 
al.\,\cite{Aljishi} have shown that the conduction tail is indeed more 
susceptible to thermal disorder than the valence tail. The temperature 
dependence can be conveniently expressed by introducing a characteristic 
energy $\rm E_0$ and fitting the electronic density of states to 
$ \rm \rho(E) \approx \exp(|E-E_f|/E_o(T))$. Aljishi et al.\,expressed 
the temperature dependence of the tail states by the slope of the 
$\rm E_o(T)$ vs. T plot and obtained a smaller value for the conduction 
band tail. We have observed a qualitative agreement of our results with 
experiment. The key observation that one should note from Fig.\,{\ref{fig3} is 
the following: the shape of the tail in the conduction band rapidly changes 
as the temperature rises from 150K to 700K. The corresponding change in the 
valence tail for the same range of energy (0.4eV) is however much less and is rather 
smooth compared to the conduction tail. Since the localized defect states 
(coming from the two dangling bonds) have been removed before plotting, this 
observation qualitatively suggests that the conduction tail states are more 
susceptible to lattice motion. It is tempting to attempt to estimate decay 
parameters for a direct comparison to experiment\cite{Aljishi}, but the sparse
sampling of tail states for this 216-atom model makes this a dangerous exercise. The
basic features do appear to be represented in our study, however.

\section{CONCLUSION}
Using accurate methods and a reasonable model of a-Si, we 
showed that there is 1) a large e-p coupling for localized 
states, 2) a significant correlation between thermal fluctuation 
of electron energy eigenvalues conjugate to localized states 
and the IPR of the model at T=0, 3) We find qualitative 
agreement with photoemission experiments\cite{Aljishi}, 4) we 
provide a simple
analytic argument for the origin of these effects. Identical 
experience with models of other amorphous materials has convinced us
that the results are correct in at least a qualitative 
way for binary glasses and amorphous materials, and perhaps other 
systems beside.
\begin{acknowledgments}
We thank the National Science Foundation for support under 
grants  DMR-0205858 and DMR-0310933. We thank Dr. J. C. Phillips
for helpful conversations and pointing out the larger significance
of the results. We thank Normand Mousseau 
for sending us the model of amorphous silicon used in this 
calculation. 

\end{acknowledgments}

\end{document}